\documentclass[a4paper,10pt]{article}

\title{Inner Product in Quantum Field Theory}
\author{Ashaq Hussain Sofi and Muhammad Ashraf Shah\\
Department of Physics, National Institute of Technology, \\ Srinagar, Kashmir-190006,  India}

\begin{document}

\maketitle

\begin{abstract}
In this paper we will analyse the inner product for a   general tensor field theory. 
We will first analyse a generalized inner product for scalar field theories. Then 
we will use it to construct a inner product for tensor field theories. We will use this inner product to 
construct the two-point function. 
\end{abstract}

\section{Introduction}
Now we also know that in nature there are four fundamental forces.
Three of those four fundamental forces are described by  gauge fields with 
compact gauge group, and is gravity, which is not a gauge 
theory with a compact gauge group. But we can also regard gravity as a gauge 
theory of diffeomorphism $\cite{5}$. In this sense all the forces of nature 
can be analyzed in the framework of gauge theory \cite{a}-\cite{b}. 
The quantization of any theory with gauge symmetry can be done using the BRST approach 
\cite{e}-\cite{smu}. 
It has been argued that
quantization of perturbative gravity in the framework of quantum field theory 
might not work since gravity is non-renormalizable . However in the
light of effective field theory there is no fundamental difference between 
renormalizable and non-renormalizable theories except the dependence on lower 
energy scale. As gravity is non-renormalizable, so  at any energy 
scale we will get new arbitrary constants for perturbative quantum gravity.
But we can use perturbative quantum gravity safely at those energies where we 
have measured all these arbitrary 
constants. With this approach it makes sense to try to look for properties of
quantum gravity by using methods of quantum field theory. To study 
the behavior of all these forces in the inflationary era  it will be essential 
to study quantum field theory on de Sitter spacetime. If the universe is 
asymptotically approaches de Sitter spacetime, then we need to study quantum field theory 
in curved spacetime also. 
Quantum field theory on curved spacetime is also important in analyzing the behavior of 
black holes. It is only by using quantum field theory on curved spacetime that Hawking radiation from 
black holes is predicted. 
Furthermore, there are many more interesting applications of quantum field theory in curved spacetime
\cite{7a}-\cite{wd1}. 
The  Unruh effect is a famous application of quantum field theory in curved spacetime. 
It   predicts that an accelerating observer will observe black-body radiation where an 
inertial observer would observe none. In other words, the background appears to be 
warm from an accelerating reference frame.  The ground state for an inertial 
observer is seen as in thermodynamic equilibrium with a non-zero temperature by the uniformly accelerated observer.
So, we need to analyse field theory on curved spacetime. However, in order to do that we 
need to derive a generalized inner product in curved spacetime. 
This can be used to construct a two-point function in curved spacetime. 
This is what we will do in this paper. 

\section{ Scalar Field Theory }
We start with a scalar field theory with the Lagrangian $\mathcal{L}$ and the action $S$. If we minimize this  action
\begin{equation}
 \delta S =  0,
\end{equation}
we will get Euler-Lagrange equation of motion as follows:
\begin{equation}
 \frac{\partial \mathcal{L} }{\partial \phi} - \nabla_c \frac{\partial\mathcal{L}}{\partial\nabla_c\phi} =0.
\end{equation}
Now if the free scalar field Lagrangian $\mathcal{L}$ given by
\begin{equation}
 \mathcal{L} = \nabla_a\phi \nabla^a \phi + m^2 \phi^2,
\end{equation}
then the classical equation of motion will given by
\begin{equation}
 (\nabla^2- m^2) \phi (x) = 0.
\end{equation}
Now we can define  a quantity called conjugate momentum current $\pi^c$ as follows:
 \begin{equation}
  \pi^c = \frac{1}{\sqrt{-g}}\frac{\partial \mathcal{L}}{\partial \nabla_c \phi}.
 \end{equation}
 Thus we have
 \begin{equation}
  \pi^c = - \nabla^c \phi.
 \end{equation}
 If $\phi_1 , \phi_2$ are two solutions of the field equations,  and $\pi^c_1 , \pi^c_2$
 the conjugate momentum currents conjugate to them, then we have
  \begin{eqnarray}
   \pi_1^c & = &- \nabla^c \phi_1, \\
   \pi_2^c & = &- \nabla^c \phi_2.
  \end{eqnarray}
  We also define a current $J^c_{(\phi_1, \phi_2)}$ as follows:
 \begin{equation}
  J^c = i [\phi_1^* \pi^c_2 - \phi_2 \pi^{*c}_1].
 \end{equation}
Now  we can write the field equations by using the definition of $\pi^c$ as
\begin{equation}
 \nabla_c \pi^c + m^2 \phi^2 = 0.
\end{equation}
Then  $\nabla_c J^c$ can be shown to vanish:
 \begin{eqnarray}
\nabla_c J^c   &=& i \nabla_c [\phi_1^* \pi^c_2 - \phi_2 \pi^{*c}_1] \nonumber\\
&=&  i[ \nabla_c \phi^*_1 \pi^c_2 - \nabla_c \phi_{2} \pi^*c_1 +\phi_1^* \nabla_c \pi^c_2 -
\phi_2 \nabla_c \pi^{*c}_1] \nonumber\\ &=&
 i[\pi^{*}_{c1} \pi^{c}_2 - \pi^{*c}_1 \pi_{2c} + m^2 (\phi^*_1 \phi_2 - \phi_1^* \phi_2 )] =  0.
 \end{eqnarray}
Thus the current $J^c$ is conserved.

We define an inner product on a space-like hyper-surface $\Sigma_c$ as follows:
 \begin{equation}
  (\phi_1, \phi_2) = \int d\Sigma_c J^c_{(\phi_1, \phi_2)}.
 \end{equation}
Let us consider the following  metric for simplicity
\begin{equation}
 ds^2 = -N^2 dt^2 + \gamma_{ij} dx^i dx^j.
\end{equation}
If we define $n_c = (N, 0)$ as the past pointing unit normal to the hyper-surface $\Sigma_c$  then, we have
\begin{equation}
 n_c n^c = g^{ab} n_a n_b \nonumber \\ =\frac{-1}{N^2} N^2 = -1.
\end{equation}
Now we can write the inner product as follows:
\begin{eqnarray}
 (\phi_1, \phi_2) &=& \int d^3 x \sqrt{\gamma} n_c J^c \nonumber \\
 &=& \int d^3 x \sqrt{\gamma}N J^0 \nonumber \\
 &=&  \int d^3 x \sqrt{-g} J^0.
\end{eqnarray}
Now note that
\begin{equation}
 \frac{d}{dt}(\phi_1, \phi_2) = \int d^3 x \partial_0 (\sqrt{-g}J^0).
\end{equation}
We have shown that  $\nabla_c J^c$ vanishes,
\begin{equation}
 \nabla_c J^c = \frac{1}{\sqrt{-g}} [\partial_0 (\sqrt{-g} J^0) + \partial_i (\sqrt{-g} J^i)] =0.
\end{equation}
We can also show by Gauss divergence theorem that
\begin{equation}
  \int d^3 x \partial_i (\sqrt{-g} J^i) = 0.
\end{equation}
So we get
\begin{equation}
\int d^4 x \partial_0 (\sqrt{-g}J^0) = 0.
\end{equation}
So this inner product does not vary with time.

Let $\phi_n$ and $\phi^*_n$ be a complete set of solutions to the field equations, 
then by definition we can expand $\phi$ as follows:
 \begin{equation}
  \phi = \sum_n [a_n \phi_n + a^*_n \phi^*_n ].
 \end{equation}
Here the sum is a shorthand notation and may contain integrals as well, for non-compact spacetime.

We also can expand $\pi^c$ in modes as follows:
\begin{equation}
 \pi^c = \sum_n [a_n \pi^c_n + a^*_n \pi^{*c}_n ].
\end{equation}
Here $\pi^c_n $  and $\pi^{c*}$ are given by
\begin{equation}
 \pi^c_n = -\nabla^c \phi_n
 \end{equation}
 and
 \begin{equation}
 \pi^{*c}_n = -\nabla^c \phi^*_{n}.
\end{equation}
So we have
\begin{equation}
  \pi^c = \sum_n  [-a_n \nabla^c \phi_n - a^*_n \nabla^c \phi^*_n ].
 \end{equation}
 We suppose
 \begin{equation}
  (\phi_n,\phi^*_m) = 0
  \end{equation}
  and
  \begin{equation}
    (\phi_n,\phi_m) = M_{nm}.
 \end{equation}

 In quantum field theory when $\phi$  is promoted to an operator $\hat{\phi}$,
 $a^*_n$ and $a_n $ become creation operators $a^{\dagger}_n$  and  annihilation operators $a_n$, respectively.

 Thus we have
 \begin{equation}
  \hat{\phi} = \sum_n [a_n \phi_n + a^{\dagger}_n \phi^*_n ].
 \end{equation}
Now as $\pi^c$ is also promoted to an operator $\hat{\pi}^c$, we also have
\begin{equation}
  \hat{\pi}^c = \sum_n  [-a_n \nabla^c \phi_n -a^{\dagger}_n \nabla^c \phi_n^*].
 \end{equation}
 Now the state $|0\rangle$ is  the state annihilated by $a_n$
 \begin{equation}
  a_n |0\rangle = 0.
 \end{equation}
This is called the vacuum state of the theory.  
Many particle states can be built  by repeated action of
$a^{\dagger}_n$ on the vacuum state. It may be noted that
as the division between $\phi$ and $\phi^*$ is not unique, 
there will be non-uniqueness in the definition of the vacuum state also $\cite{912}$.
\section{Two-Point Function}
Now the two-point function  is given by
\begin{equation}
 G(x, x') = \langle 0|\phi(x) \phi(x')|0\rangle.
\end{equation}
This can be written as
\begin{eqnarray}
 G(x, x') = \sum_{n , m}\langle 0|(a_n \phi_n +  a_n^{\dagger}\phi^*_n) (a_m\phi_m a_m^{\dagger}\phi_m|0\rangle && \nonumber \\ = \sum_{n , m}\phi_n \phi^*_m \langle 0|a_n a_m^{\dagger}|0\rangle && \nonumber \\ = \sum_{n , m}\phi_n \phi^*_m \langle 0|[a_n a_m^{\dagger}]|0\rangle &&.
\end{eqnarray}
where $[a_n, a_m^{\dagger}]$ is the commutator and thus given by
\begin{equation}
 [a_n, a_m^{\dagger}] = a_n a_m^{\dagger} - a_m^{\dagger}a_n.
\end{equation}
To calculate the two-point function  explicitly we need to calculate the effect of the commutator of the creation and annihilation operators on vacuum states. To do so we define $C_{nm}$ as follows:
\begin{equation}
 C_{nm} = \langle 0|[a_n, a_m^{\dagger}]|0\rangle.
\end{equation}
Then we have
\begin{equation}
 G(x,x') = \sum_{nm} \phi(x)_n \phi(x')^*_m C_{nm}.
\end{equation}
Now we can have
\begin{equation}
M_{nm} =  (\phi_n, \phi_m) =  \int d^3 x \sqrt{-g} J^0_{(\phi_n, \phi_m)}
\end{equation}
and
\begin{equation}
M_{mn} =  (\phi_m, \phi_n) =  \int d^3 x \sqrt{-g} J^0_{(\phi_m, \phi_n)}
\end{equation}
where
\begin{equation}
 J^0_{(\phi_n, \phi_m)} = i[\phi_n^*\pi^0_m - \phi_m \pi_n^{*0}].
\end{equation}
and
\begin{equation}
 J^0_{(\phi_m, \phi_n)} = i[\phi_m^* \pi^0_n - \phi_n \pi_m^{*0}].
\end{equation}
Now as
\begin{equation}
( i[\phi_m^* \pi^0_n - \phi_n \pi_m^{*0}])^* = i[\phi_n^*\pi^0_m - \phi_m \pi_n^{*0}],
\end{equation}
 we have
\begin{equation}
 M_{nm} = M^*_{mn}.
\end{equation}
We  also have
\begin{equation}
 [(\phi_n, \hat{\phi}),(\hat{\phi}, \phi_m)] =  \int d^3 x d^3 x' \sqrt{-g(x)} 
 \sqrt{-g(x')} [J^0_{(\phi_n, \hat{\phi})}, J^0_{(\hat{\phi}, \phi_m)}],
\end{equation}
where $J^0_{(\phi_n, \hat{\phi})}$ and $ J^0_{(\hat{\phi}, \phi_m)}$ are given by
\begin{equation}
J^0_{(\phi_n, \hat{\phi})} = i [\phi_n^* \hat{\pi}^0 - \hat{\phi} \pi_n^{*0}](t,x)
 \end{equation}
 and
\begin{equation}
  J^0_{(\hat{\phi}, \phi_m)} = i [\phi^{\dagger} \pi_m^0 -\phi_m \pi^{\dagger 0}](t, x').
\end{equation}
As $\hat{\phi}$ and $\hat{\pi}$ are hermitian, we can write
\begin{equation}
  J^0_{(\hat{\phi}, \phi_m)} = i [\hat{\phi} \pi_m^0 -\phi_m \hat{\pi}^{ 0}](t, x').
\end{equation}
Now as
\begin{equation}
 [\hat{\phi}(t,x), \hat{\pi}^ (t,x')] = i \delta (x,x')
 \end{equation}
 and
 \begin{equation}
 [\hat{\phi}(t,x), \hat{\phi}^ (t,x')]  = [\hat{\pi}(t,x), \hat{\pi}^ (t,x')] = 0,
\end{equation}
 we have
\begin{equation}
 [(\phi_n, \hat{\phi}),(\hat{\phi}, \phi_m)] = i \int d^3 x \sqrt{-g(x)} [\phi_n^*\pi_m^0 - \phi_m \pi_n^{*0}].
\end{equation}
Now, as
\begin{equation}
 i[\phi_n^*\pi_m^0 - \phi_m \pi_n^{*0}] = J^0_{(\phi_n, \phi_m)},
\end{equation}
 we get
\begin{equation}
 [(\phi_n, \hat{\phi}),(\hat{\phi}, \phi_m)] = (\phi_n \phi_m) = M_{nm}.
\end{equation}
Now we have
\begin{eqnarray}
 (\phi_n, \hat{\phi}) &=& \sum_k (\phi_n, a_k \phi_k)\nonumber\\ &=& \sum_k a_k (\phi_n , \phi_k)\nonumber \\ &=& \sum_k a_k M_{nk}
\end{eqnarray}
 and
 \begin{eqnarray}
 ( \hat{\phi}, \phi_m) &=&  [(\phi_m, \hat{\phi})]^{\dagger} \nonumber \\ &=& \sum_l [(\phi_m, a^{\dagger}_l \phi_l)]^{\dagger}\nonumber\\ &=& \sum_l a^{\dagger}_l (\phi_m , \phi_l)^* \nonumber \\ &=& \sum_l a^{\dagger}_l M^*_{ml} \nonumber \\ &=& \sum_l a^{\dagger}_l M_{lm}.
\end{eqnarray}
So we get
\begin{equation}
 [(\phi_n, \hat{\phi}),(\hat{\phi}, \phi_m)] = \sum_{kl} M_{nk}[a_k, a_l^{\dagger}]M_{lm}.
\end{equation}
Thus we can write,
\begin{equation}
 \sum_{kl} M_{nk}[a_k, a_l^{\dagger}]M_{lm} = M_{nm}.
\end{equation}
So we have,
\begin{equation}
  \sum_{kl} M_{nk}C_{kl}M_{lm} = M_{nm}.
\end{equation}
 We can write this equation  in matrix notation as
\begin{equation}
 MCM = M.
\end{equation}
So we have
\begin{equation}
 C= M^{-1}.
\end{equation}
 Now the two-point function is given by
 \begin{equation}
   G(x,x') = \sum_{nm} \psi_n \psi'_m M^{-1}_{nm}.
 \end{equation}
 \section{ Tensor Fields}
In this section we will formally generalize what we did for scalar fields to general 
non-interacting spin fields.  Let us denote the tensor field by a shorthand notation $A_{bcde...} = A_I$.
The Lagrangian for this field $A_I$ will be a scalar function of $A_I$ and $\nabla_cA_I$. 
Here again we will not consider higher derivatives as they will again lead to non-unitary
quantum field theory. In general, the  Lagrangian for  higher spin fields might be invariant 
under some gauge transformation and so we need to add some gauge fixing term.  Thus the Lagrangian 
for a general tensor field can be written as follows:
\begin{equation}
 \mathcal{L} = -\sqrt{-g}[\mathcal{L}_1  + \frac{\alpha}{2} \mathcal{L}_2],
\end{equation}
where $\mathcal{L}_1$ is the original Lagrangian and $\mathcal{L}_2$ is the contribution coming from the gauge fixing term.

 The action $S$ is given by
 \begin{equation}
  S = \int d^4 x \mathcal{L}.
 \end{equation}
If we minimize this action we get the equations of motion.

We can define $\pi^{Ic}$ here as we did in the scalar case
 \begin{equation}
  \pi^{Ic} = \frac{1}{\sqrt{-g}}\frac{\partial \mathcal{L}}{\partial \nabla_c A_I }.
 \end{equation}
Now if $A_{I1}$ and $A_{I2}$ are two solutions to the field equations then we can define the current $J^c_{(A_{1}, A_{2})}$ as follows:
 \begin{equation}
  J^c = i [A_{I1}^* \pi^{Ic}_2 - A_{I2} \pi^{*Ic}_1].
 \end{equation}
This current can again be shown to be conserved by repeating the argument for scalar field:
\begin{equation}
 \nabla_c J^c = 0.
\end{equation}
 Now we can define an inner product on a space-like hyper-surface $\Sigma_c$ as follows:
 \begin{equation}
  (A_{1}, A_{2}) = \int d\Sigma_c J^c_{(A_{1}, A_{2})}.
 \end{equation}
The inner product  here too becomes
\begin{equation}
 (A_{1}, A_{2}) =  \int d^3 x \sqrt{-g} J^0.
\end{equation}
We can  again show that this inner product does not change with time by following
a similar line of argument to what was done in the scalar case.
Now if $A_{In}$ and $A^*_{In}$ are a complete set of solutions to the classical 
equations of motion then we can expand $A_{I}$ as follows:
 \begin{equation}
  A_I = \sum_n [a_{n} A_{In} + a^*_{n} A^*_{In} ].
 \end{equation}
We also can expand $\pi^{Ic}$ in modes as follows:
\begin{equation}
 \pi^{Ic} = \sum_n [a_n \pi^{Ic}_n + a^*_n \pi^{*Ic}_n ].
\end{equation}
 We suppose
 \begin{equation}
  (A_n,A^*_m) = 0
  \end{equation}
  and
\begin{equation}
    (A_n,A_m) = M_{nm}.
\end{equation}
In quantum field theory when $A_{I}$  is promoted to an operator
$\hat{A_I}$, $a^*_n$ and $a_n $ become creation operators $a^{\dagger}_n$  
and  annihilation operators $a_n$ respectively.
 Thus we have
 \begin{equation}
  \hat{A}_I = \sum_n [a_n A_{In} + a^{\dagger}_n A^*_{In} ].
 \end{equation}
Now the two-point function  is given by
\begin{eqnarray}
 G_{II'} (x,x') &=& \langle 0| A_I(x) A_{I'}(x')|0\rangle \nonumber
 \\&=& \sum_{mn}A_{In}(x)A_{I' m}(x')\langle 0| [a_n, a_m^{\dagger}]|0\rangle.
\end{eqnarray}
If we again define $C_{nm}$ as follows:
\begin{equation}
 C_{nm} = \langle 0|[a_n, a_m^{\dagger}]|0\rangle.
\end{equation}
then following a similar line of argument to the scalar case, we can again show that
\begin{equation}
 [(A_n, \hat{A})(\hat{A},  A_m)] = M_{nm}.
\end{equation}
We can also show,
\begin{equation}
 M_{nm} = M^*_{mn}.
\end{equation}
Then we can write the above equation  in matrix notation as before
\begin{equation}
 MCM = M.
\end{equation}
So we have, just like in the scalar case
\begin{equation}
 C= M^{-1}.
\end{equation}
 Now the two-point function is given by
 \begin{equation}
   G(x,x')_{II'} = \sum_{nm} A_{In}A_{I'm} M^{-1}_{nm}.
 \end{equation}
Here the two-point function  is expected to split into two parts
\begin{equation}
   G(x,x')_{II'} = P_{II'}(x,x') + Q_{II'}(x,x').
\end{equation}
Here $P_{II'}$ and $Q_{II'}$ are contributions coming from the physical and pure gauge terms respectively.
In this paper we  constructed a inner product for a quantum field theory in curved spacetime. 

\section{Conclusion}
In this paper we analysed the quantization of quantum field theory on curved spacetime. 
We first analysed the quantization of scalar field theory. We thus constructed a inner product for 
scalar field theory. This inner product was used for constructing a two-point function. We then generalized 
these results to tensor fields. We constructed a inner product and two-point function for the most general tensor field 
theory possible. 
It was observed that the choice of vacuum state was not unique. This is because different vacuum states 
could be related to each other via a transformation. 
In fact, it is this transformation that becomes the bases of both the Hawking radiation and  Unruh effect. 
It will be interesting to analyse the  Unruh effect for tensor fields using this formalism.

\end{document}